\documentclass[preprint]{aastex}

\def\etal{{et~al. \,}}
\def\sun{_\odot}

\def\Dwa{$\,$\uppercase\expandafter{\romannumeral5}$\,$}

\def\sles{\lower2pt\hbox{$\buildrel {\scriptstyle <}
   \over {\scriptstyle\sim}$}}

\def\sgreat{\lower2pt\hbox{$\buildrel {\scriptstyle >}
   \over {\scriptstyle\sim}$}}
\def\sharpnull#1{}
\def\aa{Astron. Astrophys.\ }

\begin{document}
%-----------------------------------------------------
% private definitions
\def \m {$ M_\odot$~}
\def \l {$ L_\odot$}
\def \ro {g/cm$^{3}$}
\def \etal {\it et al.\rm}
\def \f {{\nu}}
\def \pos {{\mathbf r}}
\def \vel {{\mathbf v}}
\def \dir {{\mathbf \Omega}}
\def \dpsidt {{1 \over c} {D I \over {D t}}}
\def \dphidt {{1 \over c} {\partial \phi \over {\partial t}}}
\def \dpsidr {\mu {\partial I \over {\partial r}}}
\def \dpsidx {{d I \over {dx}}}
\def \dpsidz {\eta {\partial I \over {\partial z}}}
\def \dpsidfi {\xi {\partial I \over {r \partial \phi}}}
\def \dpsidmu {{(1-\mu^2) \over r} {\partial I \over {\partial \mu}}}
\def \divr  {{1 \over r^2} {{\partial(r^2 \mu I)} \over {\partial r}} }
\def \divx {{1 \over r} {{\partial(r \mu I)} \over {\partial r}} }
\def \divmu {{1 \over r} {\partial((1-\mu^2) I) \over {\partial \mu}} }
\def \divfi {{1 \over r} {\partial(\xi I) \over {\partial \phi}}}
\def \dt    {\Delta t}
\def \cdt   {{1 \over {c \Delta t}}} 
\def \dtau  {d\tau}

\slugcomment{\bf}
\slugcomment{Submitted to Ap.J. December 25, 2003}

\title{Two-dimensional, Time-dependent, Multi-group, Multi-angle 
Radiation Hydrodynamics Test Simulation in the Core-Collapse Supernova Context}

\author{Eli Livne\altaffilmark{1},  
Adam Burrows\altaffilmark{2}, Rolf Walder\altaffilmark{2}, 
Itamar Lichtenstadt\altaffilmark{1},
Todd A. Thompson\altaffilmark{3,4}}

\altaffiltext{1}{Racah Institute of Physics, The Hebrew University, 
Jerusalem, Israel; eli@frodo.fiz.huji.ac.il,
itamar@saba.fiz.huji.ac.il}
\altaffiltext{2}{Department of Astronomy and Steward Observatory, 
                 The University of Arizona, Tucson, AZ \ 85721;
                 burrows@as.arizona.edu, rwalder@as.arizona.edu}
\altaffiltext{3}{Astronomy Department and Theoretical Astrophysics Center, 
                 The University of California, Berkeley, CA 94720; thomp@astro.berkeley.edu}
\altaffiltext{4}{Hubble Fellow}

\begin{abstract}

We have developed a time-dependent, multi-energy-group, and 
multi-angle (S$_n$) Boltzmann transport scheme  
for radiation hydrodynamics simulations, in one and two spatial
dimensions. The implicit transport is coupled to both 1D (spherically-symmetric)
and 2D (axially-symmetric) versions of the explicit Newtonian hydrodynamics code VULCAN.
The 2D variant, VULCAN/2D, can be operated in general structured or unstructured
grids and though the code can address many problems in astrophysics it was
constructed specifically to study the core-collapse supernova problem.   
Furthermore, VULCAN/2D can simulate the radiation/hydrodynamic 
evolution of differentially rotating bodies.  We summarize 
the equations solved and methods incorporated into the algorithm and present
results of a time-dependent 2D test calculation. 
A more complete description of the algorithm is postponed to another paper.
We highlight a 2D test run that follows for 22 milliseconds the immediate
post-bounce evolution of a collapsed core.  We present the relationship between
the anisotropies of the overturning matter field and the distribution
of the corresponding flux vectors, as a function of energy group.  This is the first
2D multi-group, multi-angle, time-dependent radiation/hydro 
calculation ever performed in core collapse studies.
Though the transport module of the code is not gray and does not use flux limiters 
(however, there is a flux-limited variant of VULCAN/2D), it still does 
not include energy redistribution and most velocity-dependent terms.     

\end{abstract}


\section{Introduction}
\label{intro}

It is now clear that if stellar core collapse
simulations are constrained to be spherically-symmetric the cores 
do not supernova (Burrows, Hayes, and Fryxell 1995; Mezzacappa et al. 2001;
Rampp and Janka 2000,2002; Liebend\"orfer et al. 2001,2003; Buras 
et al. 2003; Thompson, Burrows, and Pinto 2003).
However, some calculations performed in 2D (Herant et al. 1994; Burrows, 
Hayes, and Fryxell 1995; Fryer et al. 1999)
and in 3D (Fryer and Warren 2002,2003) do explode, while others performed in 2D do not 
(Mezzacappa et al. 1998; Buras et al. 2003).  Even though a consensus is emerging that multi-D
effects are crucial to the mechanism of core-collapse supernovae, no censensus has
yet emerged concerning {\it which} multi-D effects are crucial.  The enhancement of the
efficiency of neutrino heating behind the stalled shock wave due to
neutrino-driven convection seems to play a role, but this conclusion too remains controversial.
It is suspected that the approximations to neutrino transport incorporated
into the codes which see explosions may have compromised the validity 
of those results.  Furthermore, the degree of neutrino heating in 
the semi-transparent region behind the stalled shock may depend
upon a proper treatment of multi-D transfer in the unstable
and convective core.  Moreover, the viability of doubly-diffusive
instabilites in protoneutron stars hinges upon 
neutrino transport in non-radial directions (Mayle 
and Wilson 1988; Bruenn and Dineva 1996).  Hence, 
there is now an emphasis in supernova theory circles on improving 
the treatment of neutrino radiation transport in two and three dimensions.  
We have embarked upon such a program of improvement and 
this paper is the first in a series on our evolving efforts.

In many astrophysical problems matter and radiation are strongly coupled
and are described by the system of non-linear radiation hydrodynamics equations. 
When the optical depth is large everywhere, the diffusion 
approximation can be used.  In the diffusion limit, the radiation
field is in local thermodynamic equilibrium (LTE) with the matter and
the radiation is described by the non-linear diffusion equation.
However, when important subregions of the system are
optically thin, a full Boltzmann problem must be solved. A typical example
of the latter is the core-collapse supernova problem, where the radiation field
consists of several neutrino types, the optical depth varies rapidly with time
and position, and the coupling to matter may determine the failure or success of
the neutrino-driven explosion mechanism. In that problem, the radiation field 
depends upon position, direction, frequency (energy), and species, for which any
discretization scheme becomes very large and expensive. For this reason, to date most 
time-dependent numerical radiation/hydrodynamics conducted in the astrophysical context have been limited to
one spatial dimension, or if multi-dimensional, has been either flux-limited and gray (``one-group")
(Stone, Mihalas, Norman 1992; Hayes and Norman 2003) or carried out along radial rays 
(Rampp and Janka 2000,2002; Buras et al. 2003), or both (Burrows, Hayes, and Fryxell 1995). 
The 2D calculations of Burrows, Hayes, and Fryxell (1995) incorporated 
flux-limited, gray transport along radial rays, and ignored transverse transport.
The 2D calculations of Buras et al. (2003) used a Boltzmann solver, but were  
conducted in 1D along various radial rays. 
Their hydrodynamics is, however, done in 2D.  These authors correct for transverse
radiation pressure and transverse lepton transport using 
a further update split off from the radial transport.
The error introduced by such a procedure is unknown.    
The 2D calculations of Mezzacappa et al. (1998) actually employed 
the Multi-Group-Flux-Limited Diffusion (MGFLD) neutrino heating rates
from a previous 1D simulation. These deposition rates were incorporated into a 2D hydro
simulation, but without the feedback that would have arisen from the large differences between
the 1D and 2D density and composition profiles and the shock history, 
as well as from the angular variation seen in 2D.
The ZEUS-2D radhydro code of Stone, Mihalas, and Norman (1992), and its update by
Hayes and Norman (2003), solve the zeroth- and first-moment
equations of the gray transport equation and use a short-characteristics solution of the 
static transfer equation to obtain the second moment for closure.  These codes are implementations
of the variable Eddington factor technique and avoid some of the pitfalls of flux limiters,
but since they are gray and not multi-group they are of limited utility for modern supernova simulations.  
The recent increases in computing power now make possible the
development of more sophisticated codes to perform 2D, multi-angle, multi-group,
multi-species radiation hydrodynamics simulations.  

The vast nuclear engineering literature contains numerous methods for solving 
the Boltzmann equation (Adams and Larsen 2002).  However, 
these may not be appropriate for astrophysical problems.
In astrophysics, we confront stellar configurations in which the
optical depth varies by many orders of magnitude from their centers to their peripheries
and the density and mean free paths change rapidly with radius.
In this paper, we present a test core-collapse 
simulation using a new algorithm to solve the multi-group
Boltzmann equation in one and two spatial dimensions which we incorporated into 
the broader framework of radiation hydrodynamics.  
We defer a more complete discussion of the solver, stencils, and
difference scheme to a later work (Livne et al. 2004).

We summarize aspects of our Boltzmann transport scheme in one and 
two spatial dimensions in \S\ref{differ} and \S\ref{radhydro}.
For a full radiation/hydrodynamics code, one needs to couple the radiation field to
matter through emission, absorption, scattering, and radiation pressure.  
The resulting time-, space-, energy-group-, and angle-dependent code 
has both 1D and 2D (axially-symmetric) variants.
Section \ref{radhydro} is devoted to a pr\'ecis of the radiation hydrodynamics scheme in which
the new transport module is incorporated.  We also touch on various
subtleties concerning 2D gravity and momentum conservation and the importance
of freeing the inner core to move along the symmetry axis.  In \S\ref{issues},
we summarize various computational issues concerning problem size and parallelism
philosophy.   In \S\ref{test}, we first discuss 
the results of our comparison of 1D stellar collapse simulations using
VULCAN/1D with the Feautrier/tangent-ray code ``SESAME"
\footnote{{\bf S}pherical, {\bf E}xplicit/Implicit, {\bf S}upernova,
{\bf A}lgorithm for, {\bf M}ulti-Group/Multi-Angle,
{\bf E}xplosion Simulations} (Burrows et al. 2000; Thompson, Burrows, and Pinto 2003)
and then go on to describe our preliminary 2D simulation results in the core-collapse context. 
This test simulation was carried out for 22 milliseconds of 
physical time.  We focus on the relationship between an anisotropic 
matter distribution and its corresponding anisotropic radiation field. This is the first 
2D time-dependent, multi-group, multi-angle radiation/hydro 
calculation ever performed in core collapse studies.
For these first test simulations, only electron neutrinos are 
incorporated, though the code is set up to handle the other species.
The 2D code can employ unstructured fixed or moving 
meshes and incorporates the centripetal terms due
to rotation about the axis of symmetry.   
Finally, in \S\ref{conclusion} we list the accomplishments represented by this
paper, as well as the significant work that remains yet to be done.

\section{The Transport Equations and Approach}
\label{differ}

The transport equation for the specific intensity ($I$)
\begin{equation}
 \dpsidt + \dir \cdot \nabla I + \sigma I = S    
\label{1}
\end{equation}
describes for all directions $\dir$ the additions to and subtractions from a beam of 
massless particles due to the physical processes of absorption,
scattering, emission, and streaming.  
In the notation of Morel et al. (1996), $I(\pos,\dir,\varepsilon_\nu,t)$ 
is the angular specific intensity (in energy units),
$\sigma^a(\pos,\varepsilon_\nu)$ is the absorption cross 
section times the number density (inverse absorption mean-free-path),
$\sigma^s(\pos,\varepsilon_\nu)$ is the inverse scattering mean-free-path, $\sigma$
equals $\sigma^a +\sigma^s$, $S$ (the source term)
equals $S_{em} (\pos,\varepsilon_\nu) + \sigma^s J$ (where $S_{em}$ is the emissivity), 
and $J(\pos,\varepsilon_\nu,t)$
equals ${1\over {4\pi}} \int{I d\dir}$ (the zeroth moment of $I(\pos,\dir,\varepsilon_\nu,t)$).
Scattering is currently assumed to be isotropic,
though this is not an inherent feature of our formulation, and we use the transport
cross section, $(1-\langle{\cos{\theta}}\rangle)\sigma^s_{T}$,
instead of the total scattering cross section ($\sigma^s_{T}$).
This approach has a long pedigree and has been shown to work quite well in the
1D core-collapse context (Burrows et al. 2000; Thompson, Burrows, and Pinto 2003).
In eq. (\ref{1}), an energy-group subscript for each energy $\varepsilon_\nu$
is implied.   We solve eq. (\ref{1}) implicitly and for every group and angle.  Note that the term
$\dpsidt$ is the total, Lagrangian time derivative.  In eq. (\ref{1}),
the other velocity-dependent terms, such as the Doppler (radiation compression) 
and aberration terms (Mihalas and Mihalas 1984), have been dropped.  Such velocity effects
are not dominant in the core-collapse context, but the compression
term in particular would need to be included in a future generation of supernova
simulations if neutrino viscosity effects are to be properly treated.  Coupling as they
do energy groups, angular bins, and spatial zones in the type of global
solution required for a fully implicit treatment, these two velocity-dependent
terms represent one of the most complicated numerical challenges of multi-D transport.
One component of this challenge, as yet unsolved, is to incorporate non-monotonic velocity fields, with
the resulting complicated characteristic structures, into an implicit solver.
This fact is why to date no time-dependent multi-group, multi-angle astrophysics code
has properly included them. We plan to do so in 
a later version using an operator-split, but explicit,
approach that we have developed for the spherically-symmetric code
SESAME (Thompson, Burrows, and Pinto 2003).

In spherical symmetry, the one-dimensional transport equation (eq. \ref{1}) is
\begin{equation}
 \dpsidt + \dpsidr + \dpsidmu + \sigma I = S \, .
\label{2}
\end{equation}
The discretization of the radial transport equation
is easiest when eq. (\ref{2}) is written in conservative form:
\begin{equation}
 \dpsidt + \divr + \divmu + \sigma I = S \, .
\label{3}
\end{equation}
This form expresses the fact that the transport operator ${\mathbf \Omega} \cdot
\nabla $ is the divergence of the flow in direction $\mathbf \Omega$. Straightforward integration of this
differential equation over a (small) finite volume ($2\pi r^2 dr d\mu$) provides
a conservative scheme in which appears creation and extinction
terms, along with the boundary fluxes.

In 2D axisymmetric geometry, the transport equation we solve becomes
\begin{equation}
 \dpsidt + \dpsidr + \dpsidz - \dpsidfi + \sigma I = S
\label{axi1}
\end{equation}
or in conservative form :
\begin{equation}
 \dpsidt + \divx + \dpsidz - \divfi + \sigma I = S   \, .
\label{axi2}
\end{equation}
Here, the directional cosines are just the projections of the unit vector
${\mathbf \Omega}$ (in the neutrino/photon velocity direction) on the Cartesian axes (Figure \ref{angles}):

$$ \eta=\cos(\theta) $$
$$ \mu = \sin(\theta)\cos(\phi) $$
$$ \xi = \sin(\theta)\sin(\phi) \, ,$$
where $\theta$ is the polar angle and $\phi$ is the azimuthal angle.
Generally, we distribute $\eta$
evenly from -1 to 1 and $\phi$ evenly from 0 to $\pi$, but make the number of $\phi$s
a function of $\eta$ in order to tile the
hemisphere more or less uniformly.  We need to calculate only on a hemisphere, and not the entire
sphere, because of the imposition of axial symmetry.  This is also the reason
$\phi$ is bounded by 0 and $\pi$, and by not $2\pi$.

Our numerical scheme uses two features of the transport
operator $\dir \cdot \nabla$. First, being a 
first-order differential operator along the direction
$\dir$, the transport equation can be integrated
along a characteristic that traverses the computational domain.
Second, this operator is the divergence of the velocity $\vel_{\nu}=c\dir$,
so that eq. (\ref{1}) can be regarded as a continuity equation that reflects a
conservation law for the number of particles. 
Thus, our aim was to construct a conservative difference scheme, unlike 
the method of short characteristics (e.g., van Noort, Hubeny, and Lanz 2002), that is  
motivated in part by the prevalence in astrophysics of radiation hydrodynamics problems for which 
the coupling between the radiation field and matter is stiff.
In these cases, small relative errors in the energy budget can cause 
big errors in the dynamics. The core-collapse supernova problem, in which the neutrino
radiation field contains a non-trivial fraction of the energy of the system, is a good
example of this type of physical situation. 

In VULCAN/2D, we
discretize the angular variables
using the discrete-ordinates (S$_n$) method, where the specific intensity
depends on a set of representive directions (points 
on the unit sphere) (Mezzacappa and Bruenn 1993).
Multi-group S$_n$ methods can be constructed in 
an efficient way for general grids in one, two, or three spatial dimensions. The salient 
characteristic of S$_n$ methods is the representation of the angular variable by
a set of direction cosines on the unit sphere. Unlike other
methods, such as the tangent-ray method, the parametrization of
the angular variable is independent of position and the same set of directions
applies for the entire domain. This feature makes it easy to generalize the S$_n$ method 
to any dimension and to any structured or unstructured grid.  A drawback of the S$_n$
method is its tendency at low optical depths to concentrate flux along
the rays.  However, in the core-collapse context it is the
radiation field interior to the shock wave that is most germane, and this region
is at low to modest optical depths where the anisotropy of the radiation field
is not severe.  In this work, we limit the discussion to the 1D and 2D cases only, 
since the computing resources needed for an accurate 3D solution to the 
time-dependent transport equation are currently unavailable.

\section {The Radiation Hydrodynamics Scheme: Matter/Radiation Coupling, 
Parallelization, and an Aside on Gravity}
\label{radhydro}

The numerical scheme used in VULCAN consists of a Lagrangian step,
followed by a remapping step to the Eulerian grid (Livne 1993).  This makes the
code similar in this regard to traditional ALE (Arbitrary-Eulerian-Lagrangian)
codes.  The hydrodynamical variables are all
cell-centered, except for the position and the velocity, which are node-centered.
The variables of the radiation field are also cell-centered, so that the
interaction between the radiation field and matter is properly centered.
For radiation hydrodynamics, it is convenient to define the specific intensity as the
{\it energy} specific intensity, rather than the {\it particle} specific intensity. Thus, $I_g$ is
the energy intensity in group $g$ (averaged over the group) and $I_g /{\varepsilon_\nu^g}$
is the particle intensity in group $g$ (averaged over the group), ${\varepsilon^g_\nu}$ being
the average particle (e.g., neutrino) energy of group $g$.

The time advancement in both the radiative and the 
hydrodynamical sectors is computed in the Lagrangian step,
whereas the remapping step changes only the spatial discretization of the variables
over the numerical grid. Therefore, we describe here only the
Lagrangian scheme.  The transport equation itself, for a
given source, is always computed in a fully implicit manner. 

We first advance the velocity by half a timestep:
\begin{equation}
\vel^{n+1/2}=\vel^{n}+0.5\dt(-\frac{\nabla p^n}{\rho^n} -\nabla U_G) \, ,
\label{18e}
\end{equation}
where $U_G$ is the gravitational potential (but see the end of this section).  
The position vector is then advanced using
\begin{equation}
\pos^{n+1}=\pos^{n}+\dt\vel^{n+1/2}  \, .
\label{19e}
\end{equation}
Denoting by $V$ the volume of a cell, Lagrangian mass conservation takes the form:
\begin{equation}
\rho^{n+1}=\rho^n {V^n \over V^{n+1}}     \, .
\label{20e}
\end{equation}
We then solve the adiabatic energy equation for the specific internal energy:
\begin{equation}
e^*=e^n-1/2(p^*+p^n)({1 \over \rho^{n+1}}-{1 \over \rho^{n}}) \, .
\label{21e}
\end{equation}
Equation (\ref{21e}) is iterated to convergence.
At this stage, we compute new cross sections and emission sources
\begin{equation}
S_{em}=\sum_g{\sigma^a_g J_g^{eq}}    \, ,
\label{22e}
\end{equation}
where $J_g^{eq}$ is the LTE intensity (a function of density, 
temperature, and composition).
Using those cross sections and sources, we solve the transport equation for
$I^{n+1}_g$ according to the scheme described in Livne et al. (2004).
The net change in the radiation energy density is given by:
\begin{equation}
\Delta E_r=\dt \sum_g{\sigma^a_g (J_g^{eq}-J^{n+1}_g)}  
\label{23e}
\end{equation}
and this is also minus the net change in the matter energy density. Consequently, we can compute
the final energy density, pressure, and temperature using
\begin{equation}
e^{n+1}=e^*-\Delta E_r/\rho^{n+1}                                        
\label{24e}
\end{equation}
and the equation of state. 

For the supernova problem, we also need to compute the degree of neutronization of matter
due to electron capture and other charged-current processes (Burrows and Thompson 2003). 
We obtain 
\begin{equation}
Y_e^{n+1}=Y_e^n - \dt [\sum_g{\sigma^a_g (J_g^{eq}-J^{n+1}_g) /\varepsilon^g_\nu}] {1 \over N_a \rho} \, ,
\label{25e}
\end{equation}
where $N_a$ is Avogadro's number and $Y_e$ is the elctron fraction.
Finally, we advance the velocity due to the new matter pressure by a further half timestep
and due to the radiation pressure ($\mathbf F_{rad}^{node}$) by a full timestep:  
\begin{equation}
\vel^{n+1}=\vel^{n+1/2}+0.5\dt [{1\over \rho^{n+1}}(-\nabla p^{n+1}+2\mathbf F_{rad}^{node})    
   -\nabla U_G] \, .
\label{26e}
\end{equation}
The radiation force at grid nodes is evaluated by a simple averaging process
using the radiation force at cells and the definition of the radiation flux:
\begin{equation}
{\mathbf F_{rad}^{cell} }={1 \over c}\sum_g \int{\sigma_g I_g \mathbf{\Omega} d\Omega} \, .
\label{27e}
\end{equation}

Since we are focussed here on doing radiation/hydrodynamics in two spatial
dimensions, not three, for reasonable sized grids and EOS and opacity tables we can
fit an entire hydrodynamic calculation on a single processor.  Because the 
transport solve consumes far more CPU than the hydro updating, this suggests
a path to parallelism that is both simple to implement and almost perfectly
scalable.  Given that we will handle energy redistribution explicitly, the 
energy groups are not coupled during a timestep.  Hence, we can parallelize
in energy groups (and neutrino species) and calculate the hydro on ALL processors concurrently.
Each processor handles one energy group for a given neutrino species and redundantly
calculates the hydro.  Interprocessor communication merely collects the transport
results for the neutrino energy deposition and radiation pressure calculations
that require integrals over the energy groups and angles.  The result is then
redistributed over the processors at the next timestep and the process is continued.
No domain decomposition is required.  For example, if we follow 20 energy
groups and three neutrino species, only 60 processors are required.  We use
MPI to handle the parallelization.  In practice, the communication overhead
is only 2\% to 8\% of the total run time.

Gravity is a key force in multi-dimensional astrophysical hydrodynamics.
However, many calculations in the past have employed only 
the monopole term and/or have complemented the gravitational
force term ($F_G$), written as a gradient of a potential 
in the momentum equation (eq. \ref{18e}) with a corresponding $\vec{v}\cdot F_G$
term in the energy equation.  The latter approach is perfectly reasonable, but, 
given the inherently approximate
nature of finite-difference realizations of the partial differential equations, does not 
guarantee momentum conservation nor consistency between the momentum and energy
equations when written in Eulerian form. 
To address this, we have implemented a version of the code in which the $z$-component of gravity appears
as the divergence of a stress tensor (Shu 1992; Xulu 2003).  This ensures, in principle, 
the conservation of momentum in that direction, or at least guarantees that 
in fact the gravitational force of mass parcel ``A" on ``B" is equal and opposite
to the gravitational force of mass parcel ``B" on ``A."  This 
has not been the case in past grid-based astrophysical
codes that have addressed the supernova problem.   This approach works with any 
algorithm employed to calculate the gravitational potential (we use a multipole
expansion with $\sim$20 terms) and it, along with a gridding scheme such as is
described in Ott et al. (2004), 
allows the inner core to move on the grid.  All previously published grid-based results
in supernova theory treated the inner core either in 1D (radially) or as a boundary of finite radius.
Surprisingly, no previous such publication of grid-based results allowed the core to move, something that is 
necessary if we are to address the pulsar kick problem and to treat multi-dimensional 
hydrodynamic effects with realistic gravitational feedbacks.

\section{Computational Issues of Multi-dimensional Problems}
\label{issues}

The central computational difficulty in solving the equations of radiation/hydrodynamics
in many dimensions stems from the fact that, at the very best, the number
of operations and the number of variables scales with the {\it product} 
of the number of generalized zones in the various dimensions involved. 
VULCAN/2D treats two spatial dimensions (axially-symmetric),
one time dimension, one energy-group dimension, and two angular dimensions ($\theta$ and $\phi$).
This amounts to a total of 6 dimensions.  In fact, the number of angular dimensions is 
the same as would be required in three spatial dimensions, so a full 3D
simulation would involve only one more dimension and one more hemisphere 
in momentum space (\S\ref{differ}).

In VULCAN/2D, the number of spatial cells ($N_e$) for a low-resolution
simulation might be $100\times 100 \sim 10,000$ and for a higher-resolution simulation might
be $300\times 300 \sim 100,000$.  For an S$_n$ specific intensity calculation, 
the number of $\theta$ directions might be $n=10-20$ and the number 
of $\phi$ directions might average 10, for a total of 
$N_d \sim 15\times 10 = 150$ angular bins.  The minimum number 
of energy groups required for a good core-collapse simulation 
is $N_g = 10-20$ for each neutrino species (Thompson, Burrows, and Pinto 2003).
This results in $N_e \times N_g \times N_d =$ 15,000,000 to 
300,000,000 specific intensity values at each timestep and 
for each neutrino type.  The radiation solve is always much more costly than the
hydro solve and completely dominates any run.  Doing the radiation solve for only an inner
fraction of the spatial grid has its merits and can increase speed, but we have not yet implemented this. 
Note that the cost of a corresponding 3D simulation would be larger by $\sim$100 zones$\times 2$ hemispheres.

Currently, on a standard fast processor and for the time-dependent 
test run described in \S\ref{comparison2} we required $\sim$0.5 - 2.0 minutes per energy
group per timestep.  The time-dependent run employed the equivalent of 
$16\times 10$ angular bins and $102\times 51$ spatial zones.
Since we parallelize in energy groups and neutrino species (\S\ref{radhydro})
and can expect very good scalability on most clusters, this is a benchmark number for the
current scheme. Accuracy and stability requirements currently impose 
a timestep after core bounce of $\sim$0.1-0.3 microseconds.  
Decreasing the number of angular bins ($N_d$) can increase the speed dramatically, but with
a corresponding loss of accuracy.  It is the scattering angular redistribution term
in eq. (\ref{1}) that poses the greatest numerical challenges.  A problem that
involves only absorption and emission, or that
does not involve large scattering optical depths, 
is very much simpler and faster.

\section{VULCAN Simulations}
\label{test}

\subsection{1D Comparison Test: VULCAN/1D versus SESAME}
\label{comparison1}

Though not the focus of this paper, but as a run up to the VULCAN/2D runs, we compared collapses using 
VULCAN/1D and the 1D Lagrangian radiation/hydro code SESAME,
based on Feautrier variables and the tangent-ray method, as described in Burrows et al. (2000)
and Thompson, Burrows, and Pinto (2003).
The progenitor model used for this comparison was the core of an 11 M$_{\odot}$ model
from Woosley and Weaver (1995).
The cross sections were obtained by interpolating in a three-dimensional
table in $T$, $\rho$, and $Y_e$, generated using the formulations of Burrows and Thompson (2003).
The effects of Pauli blocking by final-state electrons and 
of stimulated absorption, where appropriate, are included.
However, only electron-type neutrinos and processes were incorporated for these tests.
The equation of state we use is the finite-temperature liquid-drop model of Lattimer and Swesty (1991) 
which contains nucleons, NSE nuclei, alpha particles, photons, and electrons/positrons, 
the latter corrected for arbitrary degeneracy and relativity.  
The spherical (1D) calculations were done with the same number of radial zones (300) and energy groups (20).
The latter span an energy range up to 320 MeV.  
In the VULCAN/1D calculations, 15 S$_n$ angles were used, equally-spaced 
in $\cos(\theta)$. 

Small differences in zoning and in time stepping cause 
slight time shifts during the collapse phase
and there is a 20-millisecond shift between the two calculations
in the time of bounce (SESAME: 213 msec; VULCAN: 195 msec).
The entropies at bounce in the inner 0.5 M$\sun$ differ by 0.1 to 0.2 units.
The $Y_e$ profiles show a corresponding difference, with the values of $Y_e$ differing by 
as much as 0.01 units (inside) to 0.05 units (at $\sim$0.6 M$\sun$) 
in the inner 1.1 M$\sun$.  The outgoing deleptonization wave and the $Y_e$ trough look
quite similar and the overall lepton losses to infinity in the two calculations are within 5\%.

From the center to an interior mass of 1.0 M$\sun$, the luminosities for the VULCAN/1D and SESAME
calculations were within $\sim$1\%, but exterior to 1.1 M$\sun$ they differed by up to
$\sim$10\%.   This slight discrepancy in the outer regions can be traced to 
the inherent differences/strengths of Eulerian and Lagrangian codes, the different stencils, 
the inherent differences between tangent-ray and S$_n$ methods, and differences in the distribution of 
radial zones after bounce.  Interior to $\sim$1.25 M$\sun$, the Eddington factors for the two runs were
the same to within better than a percent, but since VULCAN/1D 
uses the S$_n$ method and S$_n$ can't resolve a strongly forward-peaked specific
intensity distribution without many, many more angular bins, in the outer 
transparent zones the Eddington factor in the VULCAN/1D run saturated
at $\sim$0.9.  The Eddington factors for the SESAME run achieved the correct value of 1.0.
However, given that the neutrino-matter energy deposition in the so-called ``gain" region occurs only in the
inner semi-transparent regions, VULCAN/1D, despite its S$_n$ 
method, provides in that critical region quite reasonable results. 
Nevertheless, updates to VULCAN that are still necessary are the energy redistribution
due to inelastic neutrino-electron scattering and the Doppler compression term.

The differences between the quantities in the 1D core-collapse simulations recently compared 
in Liebend\"orfer et al. (2003) provide a context for our 1D comparisons here.
The ORNL(Agile-Boltztran)/Garching(Vertex) codes correspond well qualitatively, and in the main quantitatively.
No evolutionary phases are importantly discrepant.  However, at a given time entropies 
differ by a few $\times$0.1, flux factors by 0.05 to 0.1, luminosities
by 10-20\%, shock radii by 10's of kilometers, and velocities by 10's of percent.  
In addition, there seem to be small numerical issues with all codes arising from
energy conservation, shock capturing, and the handling of neutrino-matter coupling, to name but a few.
Nevertheless, despite the different approaches, be they represented by SESAME, VULCAN/1D,
Agile-Boltztran, or Vertex, 1D core-collapse simulations give very similar evolutions and results.  
Note that all these codes indicate that the naively 1D (spherical) models do not explode.

\subsection{Two-dimensional, Multi-Group, Multi-Angle Simulation Results with VULCAN/2D}
\label{comparison2}

There is little in the supernova literature that addresses the issue
of aspherical neutrino radiation fields in 2D (or 3D) matter distributions.
Most of it is concerned with the effect of rotation on the flux contrast
between the pole and the equator and none of it involves the solution of
the transport or Boltzmann equation.  None of it is 
multi-group or multi-angle.  LeBlanc and Wilson (1970) performed
2D flux-limited MHD calculations to investigate the possible role
of magnetically-driven axial jets during core collapse, but did not 
address the question of the associated anisotropic radiation field.
They found that with very rapid rotation an under-energetic MHD jet punched along the poles.
Symbalisty (1984) too was concerned with the role of magnetic fields
during 2D collapse, but handled the neutrinos with a crude leakage scheme.
Janka and M\"onchmeyer (1989a,b) studied the possible ambiguity in the 
derivation of the total neutrino burst energy of SN1987a due to the
anisotropy in the neutrino flux distribution that would be caused by rapid rotation
of the protoneutron star left behind.  They derived a $\theta$  
dependence to the neutrino flux from an oblate spheroid by associating
it predominating with the solid angle subtended by the spheroid at the observer.  Pole-to-equator
contrasts of as much as three were obtained, but no multi-D
transport calculations was performed.  

Recently, Kotake, Yamada, and Sato (2003)
and Shimizu et al. (2001) used a similar approximate approach to calculate   
the flux asphericity for 2D pure-hydro rotating collapse models.  They used the code ZEUS-2D,
but without including the gray, flux-limited transport it contains.  They estimated the position
of the neutrinosphere, derived a neutrino flux enhancement at the pole, 
and concluded that such an asymmetry may play a role in the neutrino-driven mechanism
of supernova explosions and in the distribution of the debris.  We think that
there may be merit in this possibility and are planning to calculate the anisotropy
of the neutrino fluxes for rotating cores using the machinery of VULCAN/2D
\footnote{Rotation itself, and its effect on the hydrodynamic flow along the poles 
(not just the consequent enhancement of the driving neutrino fluxes), is also thought
to be a potentially important feature in the mechanism of core-collapse supernova
explosions (Burrows, Ott, and Meakin 2003).}.  

The calculations
of Fryer and Heger (2000, 2D) and Fryer and Warren (2003, 3D) are the most
complete explorations to date of the effects of rotation on the supernova mechanism.
However, these authors used a flux-limited, gray diffusion approach and did not publish
anything concerning the derived anisotropies of the radiation field, nor
on the potential role of such anisotropies in the explosion.  Their focus was mainly
on the evolution of the angular momentum and on its interaction with the convective
motions driven by neutrino heating. Finally, as mentioned in \S\ref{intro}, Buras et al. (2003)
conducted state-of-the-art 2D supernova simulations, but used multiple 1D radial
Boltzmann calculations in lieu of real 2D transport.  Our VULCAN/2D developments
are being undertaken in part to improve upon such calculations by including the second angular dimension
into full multi-group simulations performed in the 2D radiation/hydrodynamic context.

Our first 2D time-dependent radiation/hydro runs with 
VULCAN/2D, those that we publish here, are to explore aspects 
of the character of 2D transport in the core collapse context.  In a sense, they 
are intended as a ``shakedown cruise" of the code.  They are 
not meant to cover the full evolution from collapse,
through bounce, shock formation, and explosion (?). Here, we study the 
asymmetries of the radiation field using the multi-group, multi-angle code and explore the relationship
between 2D matter distributions and the corresponding radiation fields.
We also provide a first-of-its-kind glimpse at time-dependent, multi-group, multi-angle
``supernova" calculations. 

To this end, we mapped a 1D SESAME evolutionary calculation of a Woosley and 
Weaver (1995) 11 M$\sun$ progenitor core $\sim$5 milliseconds
after bounce into VULCAN/2D and continued the post-bounce calculation in full 2D for 
a further $\sim$22 milliseconds.  For this study, 
we employed only 5 $\nu_e$ energy groups, 
with group centers at 2.5, 10.8, 30, 74.6, and 177.9 MeV.  In addition, we used 
the equivalent of $16\times 10$ angular bins and $102\times 51$ spatial zones,
the latter distributed as described in Ott et al. (2004).
The calculation covered $\sim$69000 timesteps, with the timesteps themselves ranging between 
$\sim$0.1 to $\sim$0.3 microseconds.

Figures \ref{ye.velo.10} and \ref{ye.velo.22} are snapshots of the $Y_e$
distribution with velocity vectors at 10.2 and 22 milliseconds into the run.
The initial perturbations imposed have axial symmetry and are completely artificial.
These figures are 240 km$\times$240 km and 420 km$\times$420 km, respectively,
on a side.  There is a slight artifact along the poles due in part to the modest angular
resolution of the run and to the generic singularity of polar coordinates.
The positions of the stalled shock wave (roughly spherical) are
clearly seen on Figures \ref{ye.velo.10} and \ref{ye.velo.22} by the jumps in the behavior of the
corresponding velocity fields.  Inward pointing arrows in the unshocked region indicate
infall and accretion.  The slight moir\'e pattern seen in the vectors
is a graphical/visual artifact and is not in the numerical data.  
The red and off-white colors depict regions of higher $Y_e$ and proton fraction,
while the lowest $Y_e$ regions are depicted in green and blue.  The characteristic trough
in $Y_e$ between the shock and the inner core is thereby clearly indicated.
The most salient and interesting feature of the flow pattern is the pronounced vortical
motion traced by the velocity vectors between 15 and 100 kilometers radius.
Such prompt convection is expected right after core 
bounce (Burrows and Fryxell 1992; Burrows, Hayes, and Fryxell 1995).
The motion is clockwise in the positive quadrant and counterclockwise in the
negative quadrant, with a reflected counterpart in the bottom hemisphere.
The extensions of the high-$Y_e$ regions follow the velocity vectors.
A slight top/bottom asymmetry imposed in the initial model has translated into
a slight top/bottom flow asymmetry which persists, quite naturally, during the run.
In Fig. \ref{ye.velo.22}, weaker vortices are seen to be developing further out.
Due to the low spatial resolution of the grid employed,
it is unlikely that the flow field has converged and only the grossest overall
convective/overturning pattern should be considered robust.  

However, our purpose here is to determine the 2D radiation fields that result from 
aspherical matter and composition fields in time-dependent hydrodynamic supernova
flows. To depict these we first show the electron neutrino flux spectrum
vectors at 10.8 MeV.  These are provided at the same times as those 
chosen for Figs. \ref{ye.velo.10} and \ref{ye.velo.22}.
Note that 10.8 MeV is roughly at the peak of the emergent spectrum.
Figures \ref{ye.10.8.10} and \ref{ye.10.8.22} are color maps of $Y_e$ at the fiducial
times of 10.2 and 22 milliseconds, but with the flux spectrum vectors at 10.8 MeV superposed.
The shock wave is in the whitish region (high-$Y_e$) just exterior to the outer red band. 
Its positions are more easily seen in Figs. \ref{ye.velo.10} and \ref{ye.velo.22}.
Figure \ref{ent.10.8.22} depicts the 10.8-MeV flux vector distribution at 22 milliseconds,
but superposed on the corresponding entropy color map.  

As expected, at the high opacities in the inner core the fluxes are very small.
Elsewhere, the flux vectors generally point outwards, and at large distances they are more or less
radially-oriented with magnitudes that diminish as $\frac{1}{r^2}$.  However, there are many 
intriguing exceptions to this pattern.  In particular, as seen in Fig. \ref{ye.10.8.22}
and Fig. \ref{ent.10.8.22} at 22 milliseconds and between 15 and 80 kilometers (a region that brackets
the corresponding neutrinosphere) the 10.8 MeV fluxes generally point away from 
the region (red) of high $Y_e$/proton fraction and higher entropies.  In fact, at 30-40 kilometers 
(at modest to high optical depth), the flux vectors can be 45$^{\circ}$ off the radius vector.
This behavior is a consequence of the higher emissivities due to electron
capture in the $Y_e$/proton-rich, higher-entropy regime.  As Fig. \ref{ent.10.8.22} indicates,
due to the general vortical circulation pattern set up in this simulation,
the entropy distribution roughly follows the $Y_e$ distribution.  In the thick regions, the fluxes tend
(at least in part) to track the local emissivity, while at low optical depths 
in the semi-transparent region the specific intensities
from many angles ``vector" add at a given point and result in a smoothed partially radially streaming
radiation pattern.  The tendency for the radiation field at large distances to be
more smooth and spherically-distributed than the matter in the inner
regions is a generic feature of multi-D transport.
Such an effect is not easily nor accurately captured with
simpler transport schemes (e.g., Burrows, Hayes, and Fryxell 1995; 
Rampp and Janka 2002; Fryer and Warren 2002; Buras et al. 2003). 
Even so, even at 90-100 kilometers and exterior to the 10.8 MeV
neutrinosphere, the radiation field is not radial. As Fig. 
\ref{ye.10.8.22} shows, it is enhanced at angles nearest
the inner proton-rich environments near 50-60 kilometers.

One measure of anisotropy is the ratio of the radial
component of the flux to its magnitude ($\cos\theta_r =$$F_r/{\bigl|\vec{F}\bigr|}$).
Figure \ref{aniso} plots $\cos\theta_r$ versus polar angle at 10.8 MeV
and 22 milliseconds.   As Fig. \ref{aniso} shows, this ratio at 60 km
can vary by more than 40\%, while at 96 km it varies by $\sim$5\%.
The angular scale of the variation follows the angular scale of the matter 
asphericity; in this calculation at this physical time this is $\sim$20$^{\circ}$-30$^{\circ}$.
As Fig. \ref{aniso} shows, the degree of asphericity, 
as measured by $\cos\theta_r$, decreases with increasing
radius, demonstrating the ``smoothing" effect at low optical depths.  

The dependence of the flux vector and magnitude distributions on 
neutrino energy is revealed in Figs. \ref{ye.2.5.22} and \ref{ye.30.22}.
These figures are at 22 milliseconds and are similar to Fig. \ref{ye.10.8.22}, but are for
neutrino energies of 2.5 MeV and 30 MeV, respectively.  The low-energy neutrinos
are seen to assume a more radial pattern starting deeper in the core, while the
fluxes at the higher energies (and those on the tail of the emergent spectrum) are strongly
concentrated around the $Y_e$ contours and in the optically thick regions.
The latter pattern is a consequence of the higher opacities at these higher
energies. The lower opacities at 2.5 MeV cause this component to decouple
deeper in and to assume a more radial flux (and energy density) pattern at a given radius. 
Not only are the first-moment (flux) and zeroth-moment (energy density) 
anisotropies in the interior a function of neutrino
energy and radius, but the anisotropy of the emergent spectrum at infinity is a function
of energy as well.  This means that the spectrum at a given radius is generically a weak to modest function
of angle and will reflect the anisotropy of the matter due to rotation and convective overturn
interior to the shock and around the respective neutrinospheres.
For instance, at 22 milliseconds and 60 km, $\cos\theta_r$ varies
by as much as $\sim$40\% at 10.8 MeV (as Fig. \ref{aniso} indicates), 
but only by $\sim$25\% at 30 MeV and by $\sim$5\% at 2.5 MeV.
At 96 km, the corresponding numbers for 2.5, 10.8, and 30 MeV are 0.5\%, 5\%,
and 15\%, respectively, while at 125 km they are all less than 2\%.
Quite naturally, $\cos\theta_r$ goes to 1.0 at large distances.
However, another measure of anisotropy is the variation with angle 
of the magnitude of the flux itself (${\bigl|\vec{F}\bigr|}$) 
and this does not flatten at large distances.
Figure \ref{aniso2} depicts for radii from 60 to 125 km the 
dependence with polar angle of the magnitude of the flux at 10.8 MeV
and 22 milliseconds. In particular, at 125 km
${\bigl|\vec{F}\bigr|}$ for 10.8 MeV varies by as much as 80\%
from 0$^{\circ}$ to 40$^{\circ}$ polar angle and at 250 km it varies by 10-15\% from top to bottom.
At very large radii ($\sim$infinity), ${\bigl|\vec{F}\bigr|}$ at 10.8 MeV varies by $\sim$15\%
from top to bottom.  While not small, a much more anisotropic 
matter distribution, such as what rotation or
a large initial core anisotropy can generate, is required 
to produce a larger degree of flux anisotropy at infinity.
Nevertheless, the variations with angle and radius we identify here are 
important new components of supernova theory 
that previous simulations could not calculate.

Figures \ref{enerdens.10} and \ref{enerdens.22}
depict the associated maps for the energy-integrated zeroth-moment (total $\nu_e$ energy density)
at 10.2 and 22 milliseconds, respectively.  Superposed on these plots are the velocity vectors.
These distributions, and not the flux maps, are rough guides
to the distribution of neutrino heating thought to be important in the neutrino-driven
mechanism of core-collapse supernova explosions (Bethe and Wilson 1985; Burrows, Hayes, and Fryxell 1995;
Liebend\"orfer et al. 2001; Fryer and Warren 2002; Rampp and Janka 2002).
However, a more interesting plot is Fig. \ref{ed.flux.22}, which shows the relationship
at 10.8 MeV between the flux vector field and the corresponding energy-density pattern.
In this figure, the scale is 200 km$\times$200 km.  Combined in this way  
we can see more directly the correlation between a flux and an energy-density pattern.
In addition, though not obvious in Figs. \ref{ye.10.8.10} through \ref{ed.flux.22}, the ray effects
inherent in the S$_n$ method do not manifest themselves for this $N_{\theta} = 16$ 
calculation until radii of $\sim$350 kilometers,
significantly exterior to the radius of the stalled shock.

\section{Conclusions}
\label{conclusion}

We have constructed a time-dependent, multi-group, multi-angle radiation
code in two spatial dimensions and coupled it to a hydrodynamics code.  We have used the product, VULCAN/2D,
to simulate the early post-bounce evolution of a supernova core for 22 milliseconds.
The purpose of this exercise was to explore for the first time the mapping
between the aspherical, anisotropic matter distributions that obtain in
the convective protoneutron star and the corresponding radiation
field.  Highlighted were the electron neutrino flux 
and energy-density distributions and their angle dependence.
We found that these radiation distributions and their anisotropies are systematic functions 
of neutrino energy and radius and that, though the radiation field depends on the
distributions of the physical parameters, the non-local character of the solution
to the Boltzmann equation results in smoother radiation patterns in optically-thin regions.
In addition, we found that the emergent spectrum should be angle-dependent.
Even though the code was constructed to investigate core-collapse supernovae,
it can in principle be used for other astrophysical problems.  These might include 
gamma-ray bursts, accretion disk evolution, and star formation.

This paper has been a preliminary exploration and test of the new VULCAN/2D tool.
More details on the solution method and numerical approach can be found in Livne et al. (2004).
However, this study is but the beginning of a program of development
and calculation which will include the incorporation of further velocity-dependent
terms, the upgrade of the code to handle energy redistribution in an explicit manner,
full simulations with the MGFLD variant, 
simulations with rotation (cf. Ott et al. 2004),
and simulations with more energy groups and neutrino species.  Our philosophy has been to develop
in systematic fashion, and with a realistic timetable, an evolvingly sophisticated computational
capability in multi-dimensional core-collapse supernova studies, in particular, and in astrophysical
radiation hydrodynamics, in general.  Importantly, we are gearing up 
to address the neutrino-driven mechanism of core-collapse supernovae by calculating 
for much longer times, for different progenitors, and with more energy groups and neutrino species.  The
code's efficient parallelization in group and species implies that higher-energy-resolution
simulations incur little penalty.  Finally, since the code follows
transport in the angular, as well as the radial directions, 
it can explore in numerical detail and for the first 
time the viabilility of neutron-finger
or doubly-diffusive instabilities in the core-collapse 
context (Mayle and Wilson 1988).

\acknowledgments

We acknowledge discussions with  
Christian Ott, Jeremiah Murphy, Casey Meakin,
Yeliang Zhang, Salim Hariri, Ron Eastman, 
Marvin Landis, and Stan Woosley.  Support for this work is provided in part by
the Scientific Discovery through Advanced Computing (SciDAC) program
of the DOE, grant number DE-FC02-01ER41184.  T.A.T. is supported  
by NASA through Hubble Fellowship 
grant \#HST-HF-01157.01-A awarded by the Space Telescope Science
Institute, which is operated by the Association of Universities for Research in Astronomy,
Inc., for NASA, under contract NAS 5-26555. In addition, we thank
Jeff Fookson and Neal Lauver of the Steward Computer Support Group
for their invaluable help with the local Beowulf cluster and acknowledge 
the use of the NERSC/LBNL/seaborg and ORNL/CCS/cheetah machines.
Movies associated with this work can be obtained from AB, or
at http://zenith.as.arizona.edu/\~{}burrows after publication.

\clearpage
% figure 1
\figcaption{Coordinate system used for the axisymmetric transport.  The radiation
direction vector $\vec{\Omega}$ is defined in terms of $\theta$ and $\phi$,
where $\theta$ is the angle with respect to the $z$ axis at all spatial positions $z$ and $r$.
For example, $\theta = 0$, $\mu = 1$ is along the $z$ axis and $\phi = 0$ is in the $z-r$ plane.
\label{angles}}

% figure 2
\figcaption{The map of the $Y_e$ distribution at 10.2 milliseconds into the 
immediate post-bounce evolution of the core of an 11 M$_{\odot}$ progenitor from Woosley and Weaver (1995). 
Superposed are velocity vectors.  The vortical motions in this low-resolution simulation
are clearly delineated.  The scale is 240 km$\times$240 km on a side.  Green and blue
depict low $Y_e$s from $\sim$0.1 to $\sim$0.25 and red and white depict higher $Y_e$s.  The position
of the stalled shock wave is indicated by the discontinuity of the vector field.  
The slight moir\'e pattern seen in the vectors
is a graphical/visual artifact and is not in the numerical data.  See text
for details.
\label{ye.velo.10}}

% figure 3
\figcaption{Same as Fig. \ref{ye.velo.10}, but at 22 milliseconds.  The scale is 420 km$\times$420 km
on a side.  See text for details.
\label{ye.velo.22}}

% figure 4
\figcaption{At 10.2 milliseconds into the simulation, the $Y_e$ distribution using the same color
map as found in Fig. \ref{ye.velo.10}.  The scale is 240 km$\times$240 km on a side.  
The vectors denote the electron neutrino flux at 10.8 MeV 
(in ergs cm$^{-2}$ s$^{-1}$ MeV$^{-1}$, scaled for vector rendering).  
\label{ye.10.8.10}}

% figure 5
\figcaption{Same as Fig. \ref{ye.10.8.10} ($Y_e$ distribution), but at 22 milliseconds 
and using a scale of 420 km$\times$420 km on a side.  The vectors are flux vectors
at 10.8 MeV. See text for a discussion of the saliant features of this plot.
\label{ye.10.8.22}}

% figure 6
\figcaption{An entropy map of the flow at 22 milliseconds.  
The scale is 420 km$\times$420 km on a side.  Yellow denotes low entropy
regions characteristic of unshocked material, purple denotes intermediate entropy regions,
and red denotes highest entropy regions. The vectors are the fluxes 
at 10.8 MeV, as in Fig. \ref{ye.10.8.22}.
\label{ent.10.8.22}}

% figure 7
\figcaption{The ratio ($\cos\theta_r$) of the
radial component of the flux at 10.8 MeV and 22 milliseconds
to the corresponding magnitude of the flux versus
spatial polar angle from $\pi/2$ to $-\pi/2$.  This ratio
is shown for various radial distances and is one measure of the anisotropy
of the radiation field.
\label{aniso}}

% figure 8
\figcaption{Same as Fig. \ref{ye.10.8.22} at 22 milliseconds, but for 2.5 MeV electron-type neutrinos.  
The vectors trace the corresponding flux.
The scale is 420 km$\times$420 km on a side.
See text for details.
\label{ye.2.5.22}}

% figure 9
\figcaption{Same as Fig. \ref{ye.10.8.22} at 22 milliseconds, but for 30 MeV electron-type neutrinos.
The scale is 420 km$\times$420 km on a side.
See text for details.
\label{ye.30.22}}

% figure 10
\figcaption{The variation with polar angle ($\theta$) of the logarithm base 10 of 
the magnitude of the neutrino flux (${\bigl|\vec{F}\bigr|}$) at 10.8 MeV  
at 22 milliseconds into the calculation.  
The polar angle varies from $\pi/2$ to $-\pi/2$ and the flux
is in units of ergs cm$^{-2}$ s$^{-1}$ MeV$^{-1}$.  This quantity
is shown for various radial distances and is another measure of the anisotropy
of the radiation field.
\label{aniso2}}

% figure 11
\figcaption{The total energy-integrated energy density (in ergs cm$^{-3}$) of the electron-type neutrinos
at 10.2 milliseconds into the simulation.  The scale is 240 km$\times$240 km on a side.
Superposed are the corresponding velocity vectors.  White and red denote high energy densities,
and purple, blue, and green denote progressively lower energy densities.  Brown denotes
the lowest energy densities.
\label{enerdens.10}}

% figure 12
\figcaption{Same as Fig. \ref{enerdens.10}, but at 22 milliseconds and with a scale of
420 km$\times$420 km on a side.  See text for a brief discussion.
\label{enerdens.22}}

% figure 13
\figcaption{The energy density (in ergs cm$^{-3}$ MeV$^{-1}$) 
at 10.8 MeV and 22 milliseconds.  The scale is 200 km$\times$200 km.
Pink/White denotes high energy density and brown denotes low energy density.  
The sequence white, red, purple, blue, green, yellow, and brown
is a sequence of decreasing order of magnitude.  
\label{ed.flux.22}}


\begin{thebibliography}{}

\bibitem[Adams \& Larsen 2002]{A1} Adams, M.L. \& Larsen, E.W. 2002, Progress in
Nuc. Energy, 40, 3

\bibitem[Bethe \& Wilson (1985)]{bethe}
Bethe, H. \& Wilson, J.~R.~1985, \apj, 295, 14

%\bibitem[Bruenn 1985]{BR1} Bruenn, S.W. 1985, \apjs, 58, 771

\bibitem[Bruenn \& Dineva 1996]{dineva} Bruenn, S.W. \& Dineva, T. 1996, \apj, 458, L71

\bibitem[Buras et al. 2003]{buras2003} Buras, R., Rampp, M., Janka, H.-Th., \& Kifonidis, K. 2003,
Phys. Rev. Letters, 90, 1101

\bibitem[Burrows \& Fryxell 1992]{sci92} Burrows, A. \& Fryxell, B.A. 1992, Science, 258, 430

\bibitem[Burrows, Hayes, \& Fryxell 1995]{bhf_1995}
Burrows, A., Hayes, J., \& Fryxell, B.A.~1995, \apj, 450, 830

\bibitem[Burrows \etal  2000]{B2} Burrows, A., Young, T., Pinto, P., Eastman,
        R. \& Thompson, T. 2000, \apj, 539, 865

\bibitem[Burrows \& Thompson 2003]{BZ} Burrows, A. \& Thompson, T.A.  2003,
``Neutrino-Matter Interaction Rates in Supernovae: The Essential Microphysics of Core Collapse,''
to be published in {\it Core Collapse of Massive Stars},
ed. C.L. Fryer (Kluwer Academic Press).

\bibitem[Burrows, Ott, \& Meakin 2003]{bom} Burrows, A., Ott, C.D., \& Meakin, C. 2003,
to be published in the proceedings of ``3-D Signatures in Stellar Explosions:
A Workshop honoring J. Craig Wheeler's 60th birthday," held June 10-13, 2003, Austin, Texas, USA

\bibitem[Fryer et al.~(1999)]{fryer}
Fryer, C.L., Benz, W., Herant, M., \& Colgate, S. 1999, \apj, 516, 892

\bibitem[Fryer \& Heger 2000]{fryer2000} Fryer, C.L. \& Heger, A. 2000, \apj, 541, 1033

\bibitem[Fryer \& Warren 2002]{fryer2002} Fryer, C.L. \& Warren, M. 2002, \apj, 574, L65

\bibitem[Fryer \& Warren 2003]{fryer2003} Fryer, C.L. \& Warren, M. 2003, astro-ph/0309539

\bibitem[Hayes \& Norman 2003]{H1} Hayes, J.C. \& Norman, M.L. 2003, \apj, 147, 197

\bibitem[Herant et al.~(1994)]{herant}
Herant, M., Benz, W., Hix, W.R., Fryer, C.L., \& Colgate, S.A. 1994, \apj, 435, 339

%\bibitem[Hubeny 1988]{Hubeny88} Hubeny, I. 1988, Computer Physics Comm., 52, 103

%\bibitem[Hubeny 1992]{Hubeny92} Hubeny, I. 1992, in {\it The Atmospheres
%of Early-Type Stars}, ed. U. Heber \& C. J. Jeffery, Lecture
%Notes in Phys. 401, (Berlin: Springer), p. 377

%\bibitem[Hubeny \& Lanz 1995]{HubenyLanz95} Hubeny, I. \& Lanz, T. 1995,
%\apj, 439, 875

\bibitem[Janka \& M\"onchmeyer 1989a]{1monch89} Janka, H.-T. \& M\"onchmeyer 1989a, Astron. \& Astrophys., 209, L5

\bibitem[Janka \& M\"onchmeyer 1989b]{monch89} Janka, H.-T. \& M\"onchmeyer 1989b, Astron. \& Astrophys., 226, 69

\bibitem[Kotake et al. 2003]{kotake} Kotake, K, Yamada, S., \& Sato, K. 2003, \apj, 595, 304

\bibitem[Lattimer \& Swesty 1991]{Lat1} Lattimer, J.M. \& Swesty, F.D., 1991 Nucl. Phys. A, 535,331

\bibitem[LeBlanc \& Wilson 1970]{blanc} LeBlanc, J.M. \& Wilson, J.R. 1970, \apj, 161, 541

\bibitem[Liebend\"{o}rfer et al.~(2001)]{lieben2001}
Liebend\"{o}rfer, M., Mezzacappa, A., Thielemann, F.-K., Messer,
O. E. B., Hix, W.~R., \& Bruenn, S.W.~2001, PRD, 63, 103004

\bibitem[Liebend\"orfer et al. 2003]{lieben} Liebendoerfer, M. Rampp, M.,
Janka, H.-Th., \& Mezzacappa, A. 2003, astro-ph/0310662

\bibitem[Livne (1993)]{livne:93}
Livne, E. 1993, \apj, 412, 634

\bibitem[Livne et al. 2004]{livne04} Livne, E., Lichtenstadt, I., Burrows, A., Walder, R. 2004, in preparation.

\bibitem[Mayle \& Wilson 1988]{mayle} Mayle, R. \& Wilson, J. R. 1988, \apj, 334, 909

\bibitem[Mezzacappa \& Bruenn (1993)]{mezz93B}
Mezzacappa, A. \& Bruenn, S.W.~1993, \apj, 410, 669

\bibitem[Mezzacappa et al. 1998]{mezz98} Mezzacappa, A., Calder, A.C, Bruenn, S.W., Blondin, J.M.,
 Guidry, M.W., Strayer, M.R., Umar, A.S. 1998, Ap.J., 495, 911

\bibitem[Mezzacappa et al.~(2001)]{mezz2001}
Mezzacappa, A., Liebend\"{o}rfer, M., Messer, O.E.B.,
Hix, W.R., Thielemann, F.-K., \& Bruenn, S.W.~2001, PRL, 86, 1935

\bibitem[Mihalas (1984)]{mihalas84}
Mihalas, D. \& Mihalas, B., {\it Foundations of Radiation Hydrodynamics},
New York, Oxford University Press, 1984

\bibitem[Morel \etal 1996]{M1} Morel, J.E., Wareing, T.A., \& Smith, K. 1996, J.Comput.Phys., 128, 445

\bibitem[Ott et al. 2004]{ott} Ott, C.D., Burrows, A., Livne, E., \& Walder, R. 2003, accepted to
\apj, v. 600, January 2004

\bibitem[Rampp \& Janka (2000)]{rampp2000}
Rampp, M. \& Janka, H.-Th. 2000, \apjs, 539, 33

\bibitem[Rampp \& Janka (2002)]{rampp20022}
Rampp, M. \& Janka, H.-Th. 2002, \aa, 396, 331 

\bibitem[Shimizu et al. 2001]{shimizu} Shimizu, T., Ebisuzaki, T., Sato, K., \& Yamada, S. 2001, \apj, 552, 756

\bibitem[Shu 1992]{shu92} Shu, F.H., 1992, ``The Physics of Astrophysics," Volume II 
(University Science Books: Mill Valley, CA), p. 46.

\bibitem[Stone, Mihalas, \& Norman 1992]{stone} Stone, J.M., Mihalas, D., \& Norman, M.L. 1992,
Ap.J. Suppl., 80, 819

\bibitem[Symbalisty 1984]{symbal} Symbalisty, E.M.D. 1984, \apj, 285, 729 % His Figure 9: Iso \nu-energy density

\bibitem[Thompson \etal 2003] {Tod1} Thompson, T.A., Burrows, A., \& Pinto, P.A., 2003, \apj, 592, 434

\bibitem[van Noort \etal 2002] {hub2} van Noort, M., Hubeny,I. \&
Lanz, T. 2002, \apj, 568, 1066

\bibitem[Wareing \etal 2001]{W1} Wareing, T.A., McGhee, J.M., Morel, J.E. \&
Pautz, S.D 2001, Nuc. Sci. \& Eng., 138, 256

\bibitem[Woosley \& Weaver (1995)]{woosley_weaver}
Woosley, S.E. \& Weaver, T.A. 1995, \apjs, 101, 181

\bibitem[Xulu 2003]{xulu} Xulu, S.S. 2003, hep-th/0308070

\end{thebibliography}
\end{document}